\documentclass{bioinfo}
\copyrightyear{2005}
\pubyear{2005}
\usepackage{url}

\begin{document}
\firstpage{1}

\title[Fixed-Length Coding DNA Compression]{A Fixed-Length Coding Algorithm for DNA Sequence Compression(Draft,using {\it Bioinformatics} \LaTeX \ template)}
\author[Jie Liu \textit{et~al}]{Jie Liu\,$^{\rm a}$, Sheng Bao\,$^{\rm b}$\footnote{to whom correspondence should be addressed},Zhiqiang Jing\,$^{\rm c}$,Shi Chen\,$^{\rm c}$}
\address{$^{\rm a}$Dept. of Computer Science and Technology,Nanjing Univ.of P \& T,Nanjing,Jiangsu 210046,CHINA, $^{\rm b}$Dept. of Information Engineering,Nanjing Univ.of P \& T,Nanjing,Jiangsu 210046,CHINA, \\ $^{\rm c}$School of Life Science,Nanjing
University,Nanjing,Jiangsu 210093,CHINA}
\maketitle

\begin{abstract}

\section{Summary:}
While achieving a compression ratio of 2.0 bits$\slash$base ,the new
algorithm codes non-N bases\footnote{``non-N bases" refers to bases excpet N.Thus A,T,G or C. \ N stands for unknown base.} in fixed length.It dramatically reduces the time
of coding and decoding than previous DNA compression algorithms and some
universal compression programs.

\section{Availability:}http://grandlab.cer.net/topic.php?TopicID=50

\section{Contact:} \href {name@host.com}{forrestbao@gmail.com} {glascholar@263.net} {jzq8255@sina.com}
\end{abstract}

\section{Introduction}
File compression reduces file redundancy in order to represent more information in less signs in accordance with information theory \citep{Shannon1948}.As specified algorithm for image,audio and video are devised,it is necessary to devise the algorithm specified for DNA compression since huge amounts of DNA sequences needs to be stored and communicated to a large number of people.\citep{ACM2005} \citep{CS_BS}Although some universal compressors\citep{LZ77} are used in bioinformatics field,new DNA sequence compressors are being devised,such as Biocompress\citep{Biocompress} ,Biocompress-2 \citep{Biocompress2}, GenCompress \citep{GenCompress},Cfact \citep{Cfact}, DNACompress\citep{Repeat},CTW-LZ \citep{CTWLZ} and GeNML \citep{ACM2005}.

But they have a big problem,too slow execution.We improve our
LUT\citep{LUT} and use new file structure to identify different types of segment.The most advantage of this algorithm is fast
execution and easy implementation.The compression and decompression
speed is much faster than many newly-devised DNA-specified and well-known universal compression
algorithms.Since the
compression ratio is not much higher than existing ones and the compression
speed is impressively fast,our algorithm is an applicable algorithm
for fast DNA sequence compression,especially for database records compression.

\begin{methods}
\section{methods}
\subsection{Coding non-N bases}
non-N bases have four prossibilities:A,T,G or C.Each of them corresponds to a unique combination of two binary numbers.
We code them as A to 00,T to 01,G to 10,C to 11.Thus,we take 1 Byte(8 bits)to store 4 bases.
\subsection{File format of compressed file}
We will begin discussing file structure with the definition of ``section",a DNA segment.``section" contains a serie of successive Ns and ends at the last non-N base ahead the next serie of successive Ns.
``section" is the basic element to which we consider in compression and decompression.

Each DNA section corresponds to a ``file section" which contains the information of both N and non-N bases in this section.
Each file section starts with an 8 Bytes head.The first 4 Bytes records the amount of N bases whereas the following 4 Bytes records the number of non-N bases in this section.This means that each section corresponds to a real DNA segment which has at most $2^{32}$ N bases and $2^{32}$ non-N bases respectively.

The coded values of non-N bases locate after the head.The coded information is written into destination file Byte by Byte. Considering the number of non-N bases in a section may not be a multiple of 4,the second 4 Bytes in head provides accordance for decompression program about how many bit values are effective and where the next section begins.
\subsection{Compression algorithm}
The compression program reads characters from source file and writes coded binary values into destination file,restricted by the file format defined above.
Steps of compression algorithm is as below.
\begin{enumerate}
\item Preserve 8 Bytes at the beginning of file section.
\item Count the number of Ns in a successive N bases segment(To a sequence starts from non-N bases,this value is 0.) until the first non-N base is encountered.Write the number of Ns into the first 4 Bytes of the section head.
\item Code all following successive non-N bases into destination file while count their number until the next N is encountered.Write the number of non-N bases into the second 4 Bytes of the section head.
\item Move the file writing pointer to the beginning of next Byte in destination file.
\item Repeat all the above until the end-of-file is encounted.
\end{enumerate}
\subsection{Decompression algorithm}
Steps of decompression algorithm is as below.
\begin{enumerate}
\item Read the head(the first 8 Bytes) to obtain information about how many Ns are in this section and how many non-N bases are effective.
\item Write Ns into destination file,the decompressed file,according to the number written in the first 4 Bytes of the head.
\item Read the following 4 Bytes to determind how many bits should be decoded then and where the next section begins.The next section begins from the most nearby next Byte of compressed file.
\item Decode effective bits whose amount is recorded in last 4 Bytes of this section's head.Move reading pointer to the next section.
\item Repeat all the above from the beginning of the next section until the end-of-file is encounted.
\end{enumerate}
\subsection{Algorithm implementation}
The C++ and C source codes of algorithm implementation are avaliable at the website provided in Abstract of this paper.
\end{methods}

\section{experiments}
\label{experiment}
Experiments are operated to test our algorithm. Codes for testing the algorithm are
continually revising.\citep{OurCode}These tests are performed on a computer whose CPU is AMD Duron 750MHz and operating system is
MagicLinux 1.2 (Linux Kernel 2.6.9) without swap partition. Testing programs are
executed at multiuser text mode and compiled by gcc 3.3.2 with optimization level O3.
The file system is ext3.Files are stored on a 4.3 GB Quantum Fireball hard disk with 5400 RPM.
Table \ref{size} compares compression ratio while table \ref{time} compares running time.

\begin{table}[!ht]
\processtable{Comparison on compression ratio \label{size}}
{\begin{tabular}{lccccc}\toprule
sequence 	& size 		& ours 		& DNA 		& Gzip 		&bzip2\\\midrule
atatsgs 	& 9647 		& 2.0068	& -- 		& 2.1702 	& 2.15\\
atef1a23 	& 6022 		& 2.0113	& -- 		& 2.0379  	& 2.15\\
atrdnaf 	& 10014 	& 2.0068	& -- 		& 2.2784   	& 2.15\\
atrdnai 	& 5287 		& 2.0125	& -- 		& 1.8846   	& 1.96\\
chmpxx 		& 121024 	& 2.0005	& 1.6716 	& 2.2821   	& 2.12\\
chntxx 		& 155939 	& 2.0004	& 1.6127 	& 2.3349   	& 2.18\\
hehcmvcg 	& 229354 	& 2.0003	& 1.8492 	& 2.3278   	& 2.17\\
hsg6pdgen	& 52173 	& 2.0013	& --		& 2.2444   	& 2.07\\
humdystrop	& 38770 	& 2.0018	& 1.9116 	& 2.3633   	& 2.18\\
humghcsa 	& 66495		& 2.0010	& 1.0272 	& 2.0655   	& 1.31\\
humhdabcd	& 58864 	& 2.0011	& 1.7951 	& 2.2399   	& 2.07\\
humhprtb 	& 56737 	& 2.0012	& 1.8165 	& 2.2670   	& 2.09\\
mmzp3g 		& 10833 	& 2.0065	& -- 		& 2.3225   	& 2.13\\
mpomtcg 	& 186609 	& 2.0004	& 1.8920 	& 2.3291   	& 2.17\\
mtpacg 		& 100314 	& 2.0007	& --	 	& 2.2922   	& 2.12\\
vaccg 		& 191737 	& 2.0004	& 1.7580 	& 2.2520   	& 2.09\\
xlxfg512 	& 19338 	& 2.0035	& -- 		& 1.8310   	& 1.80\\
chr10(rice)   & 22432531        & 2.0000	& --		& 2.4498	& 2.3033\\
Average 	& -- 		& 2.0031	& 1.7037 	& 2.3224   	& 2.0674\\ \botrule
\end{tabular}}{Compress ratio of other algorithms are cited from their original papers.As the compression ratio of newly-devised algorithms are similiar,we take DNACompress as an example."ours" refers to our algorithm.DNA stands for DNACompress.The unit of file size is bit rather than Byte.}
\end{table}

\begin{table}[!ht]
\processtable{Comparison on running time \label{time}}
{\begin{tabular}{lccccc}\toprule
sequence	& Gzip(s)&encode(CLK)	& decode(CLK)  	& encode(s)	& decode(s)\\ \midrule
atatsgs 	& 0.013 & $<$10000	& $<$10000 	& $<$0.01	& $<$0.01\\
atef1a23 	& 0.011	& $<$10000 	& $<$10000 	& $<$0.01 	& $<$0.01\\
atrdnaf 	& 0.014	& $<$10000 	& $<$10000 	& $<$0.01   	& $<$0.01\\
atrdnai 	& 0.010	& $<$10000 	& $<$10000 	& $<$0.01    	& $<$0.01\\
chmpxx 		& 0.105	& 10000 	& 10000 	& 0.01		& 0.01   \\
chntxx 		& 0.135 & 20000 	& 20000 	& 0.02    	& 0.02\\
hehcmvcg 	& 0.198 & 30000 	& 30000 	& 0.03    	& 0.03\\
hsg6pdgen	& 0.044	& $<$10000 	& $<$10000 	& $<$0.01 	& $<$0.01\\
humdystrop	& 0.037	& $<$10000 	& $<$10000 	& $<$0.01	& $<$0.01\\
humhdabcd 	& 0.050 & $<$10000 	& $<$10000 	& $<$0.01	& $<$0.01\\
humghcsa 	& 0.055	& 10000 	& 10000 	& 0.01    	& 0.01\\
humhprtb 	& 0.049	& $<$10000 	& $<$10000 	& 0.01    	& 0.01\\
mmzp3g 		& 0.014	& $<$10000 	& $<$10000 	& $<$0.01    	& $<$0.01\\
mpomtcg 	& 0.100	& 20000 	& 30000 	& 0.02    	& 0.03\\
mtpacg 		& 0.088	& 10000		& 10000 	& 0.01    	& 0.01\\
vaccg 		& 0.164	& 30000 	& 20000 	& 0.03    	& 0.02\\
xlxfg512 	& 0.018	& $<$10000 	& $<$10000 	& $<$0.01    	& $<$0.01\\
chr10(rice)  	& 9.5 & 3460000 	& 3510000 	& 3.46    	& 3.51\\ \botrule
\end{tabular}}{``Gzip" includes the total of time elapsed in both
compression and decompression by Gzip.More experiments indicate that bzip2 takes more time to perform same operation.Following four fields list the time elapsed in compression and decompression respectively.``encode" means compression while ``decode" means decompression.Each operation is evaluated in two units,CPU clock and second.}
\end{table}


\section{Discussion}
The performance of a compression algorithm has two sides,the compression ratio and the running time.Many newly-devised DNA compression algorithms focus on compression ratio while ignore the running time.But the time occupation of obtaining a little lower compression ratio is very high.
Many of them run 100 times slower than universal compression algorithm,according to Table 2 of Chen's paper \citep{Repeat}.Our algorithm runs many times faster than Gzip which is 100 times faster than newly-devised algorithms.Considering the compression ratio and running time both advance traditional compressors(Gzip and bzip2) considerablly,our algorithm is a wise choice of replacing them.It is more useful in  those fields which need fast running,such as database.


\begin{thebibliography}{}
\bibitem[Bao {\it et~al}.,  2005]{LUT}Bao,S., Chen,S., Jing,Z.-Q., Ren,R. A DNA Sequence Compression Algorithm Based on LUT and LZ77(2005) [Online] \url{http://arxiv.org/cs/0504100}

\bibitem[Chen {\it et~al}., 2001]{GenCompress}Chen,X.,Kwong,S. and Li,M. (2001) A compression algorithm for DNA sequences {\it IEEE Engineering in Medicine and Biology} {\bf 20},61-66

\bibitem[Chen {\it et~al}.,  2002]{Repeat}Chen,X., Li,M., Ma,B. and Tromp,J.(2002) DNAcompress:fast and effective dna sequence compression, {\it Bioinformatics}, {\bf 18},1696-1698

\bibitem[Cohen, 2005]{CS_BS}Cohen,J.(2005) Computer science and bioinformatics, {\it COMMUNICATIONS OF THE ACM}, {\bf 48},72-78.

\bibitem[Grumbach {\it et~al}., 1993]{Biocompress}Grumbach,S. and Tahi,F. (1993) Compression of DNA sequences.In {\it Proceedings of the Data Compression Conference},340-350

\bibitem[Grumbach {\it et~al}., 1994]{Biocompress2}Grumbach,S. and Tahi,F. (1994) A new challenge for compression algorithms:Genetic sequences. {\it J. Inform.Process.Manage} {\bf 30},875-886

\bibitem[Korodi {\it et~al}.,  2005]{ACM2005}Korodi,G. and TABUS,I.(2005) An efficient normalized maximum likelihood algorithm for DNA sequence compression,{\it ACM Transactions on Information Systems}, {\bf 23}, 3-34

\bibitem[Liu {\it et~al}., 2005]{OurCode} Liu,J. and Bao,S. (2005) C and C++ implementaion of algorithm and binary executable files for Linux [Online] \url{http://grandlab.cer.net/topic.php?TopicID=50}

\bibitem[Matsumoto {\it et~al}., 2000]{CTWLZ} Matsumoto,T.,Sadakane,K. and Imai,H. (2000) Biological sequence compression algorithms {\it Genome Informatics Workshop} Universal Academy Press ,43-52

\bibitem[Rivals {\it et~al}., 2000]{Cfact} Rivals, E., Delahaye, J.,Dauchet, M., AND Delgrange,O. (1995) A guaranteed compression scheme
for repetitive DNA sequences. {\it Tech. Rep. IT–285, LIFL Lille I Univ. }

\bibitem[Shannon, 1948]{Shannon1948}Shannon,C.E. A mathematical theory of communication(1948) {\it The Bell System Technical Journal}, {\bf 27}, 379-423,623-656

\bibitem[Ziv {\it et~al}.,  1977]{LZ77}Ziv,J. and Lempel,A.  A Universal Algorithm for Sequential Data Compression(1977) {\it IEEE Transactions on Information Theory},{\bf 23},337-343

\end{thebibliography}
\end{document}